\documentclass[a4paper,fleqn,usenatbib]{mnras}

\usepackage{newtxtext,newtxmath}

\usepackage[T1]{fontenc}
\usepackage{ae,aecompl}

\usepackage{graphicx}	
\usepackage{amsmath}	
\usepackage{amssymb}	
\usepackage{refcount}

\newcommand{\mytilde}{\raise.17ex\hbox{$\scriptstyle\mathtt{\sim}$}}

\title[Late-time observations of Swift J1112.2-8238]{Late-time observations of the relativistic tidal disruption flare candidate Swift J1112.2-8238}

\author[G.C. Brown et al.]{
G.C.~Brown$^{1}$\thanks{E-mail: g.c.brown@warwick.ac.uk},
A.J.~Levan$^1$, E.R.~Stanway$^1$, T.~Kr{\"u}hler$^{2}$,  N.R.~Tanvir$^3$,\newauthor L.J.M.~Davies$^4$, A.~Fruchter$^5$, S.B.~Cenko$^6$, B.D.~Metzger$^{7}\,^{8}$\\
  $^1$Department of Physics, University of Warwick, Gibbet Hill Road, Coventry CV4 7AL, UK\\
  $^2$Max-Planck-Institut f\"{u}r extraterrestrische Physik, Giessenbachstra\ss e, 85748 Garching, Germany \\
  $^3$Department of Physics and Astronomy, University of Leicester, University Road, Leicester, LE1 7RH, UK \\
  $^4$ICRAR, University of Western Australia, 7 Fairway, Crawley, WA 6009, Australia	\\
  $^5$Space Telescope Science Institute, 3700 San Martin Drive, Baltimore, MD 21218, USA \\
  $^6$Astrophysics Science Division, NASA Goddard Space Flight Center, Mail Code 661, Greenbelt, MD 20771, USA \\
  $^7$Columbia Astrophysics Laboratory, Pupin Hall, New York, NY, 10027, USA \\ 
  $^8$Joint Space-Science Institute, University of Maryland, College Park, MD 20742, USA \\ 
}

\date{Accepted XXX. Received YYY; in original form ZZZ}

\pubyear{2016}

\begin{document}
\label{firstpage}
\pagerange{\pageref{firstpage}--\pageref{lastpage}}
\maketitle

\begin{abstract}
We present late-time follow-up of the relativistic tidal disruption flare candidate Swift J1112.2-8238. We confirm the previously determined redshift of $z=0.8900\pm0.0005$ based on multiple emission line detections. {\em HST} imaging of the host galaxy indicates a complex and distorted morphology with at least two spatially distinct components. These are offset in velocity space by less than 350\,km\,s$^{-1}$ in VLT/X-Shooter observations, suggesting that the host is undergoing interaction with another galaxy. The transient position is consistent to 2.2$\sigma$ with the centre of a bulge-like component at a distance of 1.1$\pm$0.5\,kpc from its centre. Luminous, likely variable radio emission has also been observed, strengthening the similarities between Swift J1112.2-8238 and other previously identified relativistic tidal disruption flares. While the transient location is $\mytilde2\sigma$ from the host centroid, the disrupted nature of the host may provide an explanation for this. The tidal disruption model remains a good description for these events.

\end{abstract}




\section{Introduction}\label{sec:int}

A growing number of detected candidates \citep[e.g.][]{Holoien2014,Arcavi2014,Holoien2016a} makes the study of tidal disruption flares (TDFs) a rising field in transient astronomy. These rare events represent the disruption and subsequent accretion of stars by supermassive black holes \citep[SMBHs;][]{Rees1988}. Their detection and study provides useful insights into accretion onto SMBHs, the same process that powers active galactic nuclei, but over human timescales of months to years. Analysis of the population of TDFs may provide a new way to determine the mass and spin population of supermassive black holes in galaxies that are inactive and too distant for other methods to be viable.

In the last few years, a possible new population of transient events has been discovered. As opposed to the typically UV, optical and soft X-ray bright ``thermal'' TDFs \citep[see][for a review]{Komossa2015}, two events have been characterised by intense $\gamma$-ray and hard X-ray emission dominated by power law spectral components that remain bright ($\mytilde10^{47}$\,erg\,s$^{-1}$) and highly variable for well in excess of $10^6$ seconds \citep[][]{Levan2011,Burrows2011,Cenko2012,Pasham2015}. This is accompanied by more moderate optical/NIR emission ($\mytilde10^{42-43}$\,erg\,s$^{-1}$) and luminous radio flares with moderate inferred Lorentz factors \citep[][]{Zauderer2011,Berger2012}. Each candidate has also been found to be associated with the centres of dwarf, star-forming galaxies \citep[][]{Levan2011,Cenko2012,Pasham2015}.

While few in number, these events have been considered to be a separate class. The most popular explanation is that these are examples of a subclass of TDF that also launches a moderately relativistic jet which enhances the observed emission through collimation and relativistic beaming, in a way analogous to both blazars and on-axis GRBs, making them feasible for detection with the {\em Swift} Burst Alert Telescope ({\em Swift}-BAT). With considerably higher apparent luminosities, these relativistic TDFs (rTDFs) are observable at much higher redshifts than their relatively local thermal cousins, with the first two events of the class, Swift J164449.3+573451 \citep[henceforth Swift J1644+57;][]{Levan2011,Bloom2011,Zauderer2011} and Swift J2058.2+0516 \citep[Swift J2058+05;][]{Cenko2012,Pasham2015} being detected at redshifts of $z=0.353$ and $z=1.185$ respectively. It is unclear as yet whether these events represent a truly separate sub-class, with members possessing a relativistic jet absent in thermal TDFs, or whether they are merely the extreme end of a continuum with varying strengths of jet.

Radio observations of thermal TDFs have, for the most part, resulted in non-detections, placing a tight constraint on any putative jet energies \citep{vanVelzen2013,Bower2013}. However in a few rare cases there has been evidence for these thermal sources producing low-level radio jets. TDF candidate ASSASN-14li \citep[][]{Holoien2016a} was observed to be X-ray bright, albeit at several orders of magnitude below that in the case of Swift J1644+57, and also detected in radio emission with a comparable flux ratio \citep{VanVelzen2015, Alexander2016}.  Similarly XMMSL1 J0740‑85 was both X-ray and radio luminous with a comparable luminosity to ASSASN-14li \citep[][]{Saxton2016,Alexander2016}. Given that late time radio emission is believed to be largely isotropic \citep[][]{Genezerov2017}, this may suggest a range of jet energies powering emission in these events. In order to determine this, further examples of both thermal and relativistic TDFs must be found and studied in detail.

In \citet[][henceforth B15]{Brown2015} it was shown that the properties of Swift J1112.2-8238 (Swift J1112-8238) were consistent with those of the previous rTDF candidates. Observations indicated that the event had an extragalactic origin, being associated with an extended source that exhibited a single emission line in optical spectroscopy. This was interpreted as the [O{\sc ii}]$\lambda$3727\,\AA\ emission doublet, placing the host at a redshift of $z=0.89$. At this redshift, the inferred properties of the flare are consistent with the previous relativistic tidal disruption flare candidates in luminosity, evolution and spectral energy distribution, and thus it represents the third member of this class

However the redshift could not be confirmed with the available data as the doublet nature of the line was not resolved and no other emission lines were present. Further, a key diagnostic of the previous rTDF candidates had not yet been observed, namely the rising radio flare associated with the presumed relativistic jet. There was also a need to strengthen the constraints on the position of the flare within its host in order to confirm the transient's nuclear origin, the available optical imaging being of too low resolution to resolve structure within the host. Here we present further observations including high resolution optical imaging from the {\em Hubble Space Telescope} ({\em HST}), medium resolution spectroscopy from X-Shooter on the VLT, and radio observations with the Australia Telescope Compact Array, between them capable of answering many of the unresolved questions surrounding this source.

All magnitudes presented in this paper are in the AB magnitude system.  Where necessary, we use a standard $\Lambda$CDM cosmology with $H_0$ = 70\,km\,s$^{-1}$\,Mpc$^{-1}$, $\Omega_M = 0.3$ and $\Omega_\Lambda = 0.7$.

\section{Observations}\label{sec:obs}

\subsection{HST Imaging}\label{HST}

Observations of the host of Swift J1112-8238 were obtained with Wide Field Camera 3 \citep[WFC3,][]{Dressel2016} on the {\em Hubble Space Telescope} on 2015 March 7 (MJD 57088), $\mytilde4$ years after the TDF event. Images were obtained in the F160W filter ($\mytilde\,H$-band) beginning at 19:39{\sc UT} for an exposure time of 997\,s ($4\times249$) and the F606W ($\mytilde\,V$-band) filter beginning at 20:21{\sc UT} with an exposure time of 1568\,s ($4\times392$).

For the optical observations, the target was placed near the lower-left corner of the CCD in order to reduce the inherent charge transfer efficiency (CTE) issues of the WFC3 UVIS chip. The images were CTE-corrected via the method of \citet[][]{Anderson2010}. The position angle of the observations was chosen to ensure that diffraction spikes from a nearby bright (R = 15.8 mag) star would not interfere with the target. Each set of exposures was obtained with a sub-pixel dither pattern allowing the images to be redrizzled to half the native pixel scale (resulting in pixel scales of 0.065$^{\prime\prime}$/pix and 0.02$^{\prime\prime}$/pix in the F160W and F606W images respectively) and combined using the {\sc pyraf} routine {\sc astrodrizzle} \citep[][]{Fruchter2010}.

\begin{figure}
\begin{center}
\includegraphics[width=6.8cm]{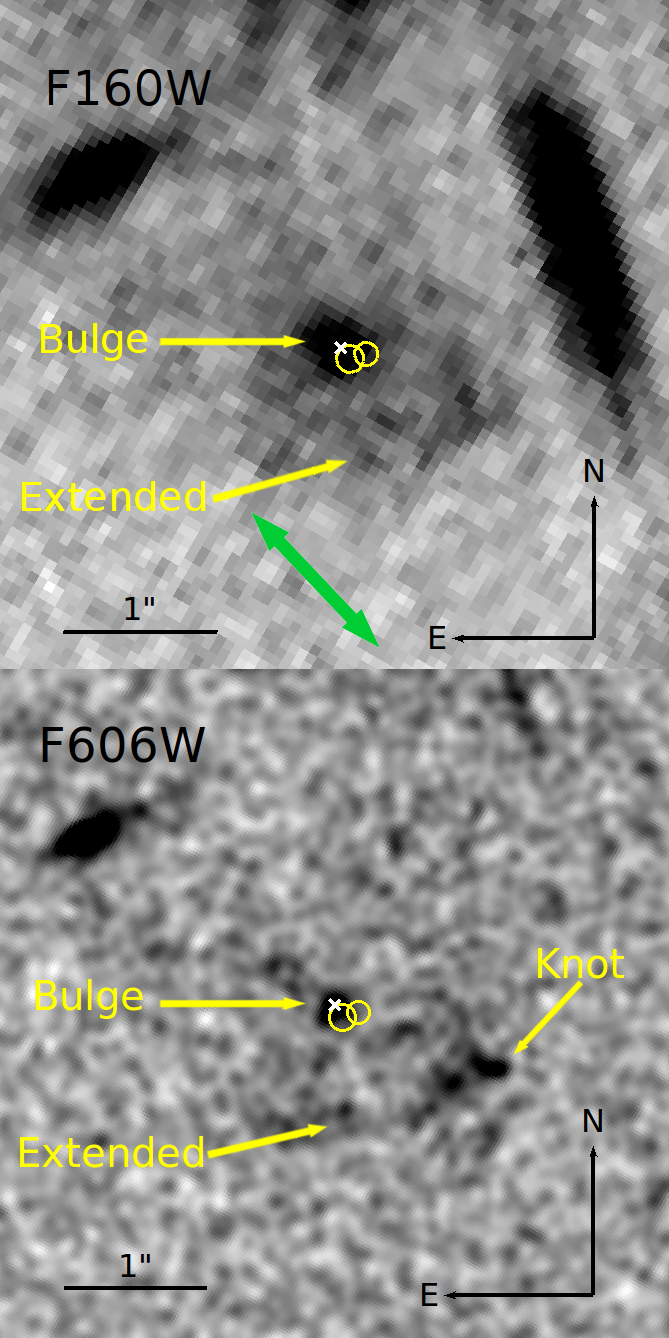}
\caption[{\em HST} WFC3 images of the host of Swift J1112-8238 in the F160W and F606W bands]{{\em HST} WFC3 images of the host of Swift J1112-8238 in the wavebands (Upper) F160W and (Lower) F606W. The F606W imaging is presented with a 3 pixel smoothing applied. The complex morphology hinted at in the lower resolution GMOS imaging (discussed in B15) is clear in this higher resolution {\em HST} imaging. The ``host'' as determined in B15 appears to be a combination of two separate components, a bulge-like feature and a more extended component (indicated on the images). Also note the existence of a ``knot'' of emission within the extended component on the western edge of the host complex, most apparent in the F606W imaging. The complex morphology of the host could be indicative of a bulge embedded in a wide disk with a small region of strong star formation. However the asymmetry of the morphology points instead to an ongoing merger or interaction of two galaxies. The position of the transient as determined in the two early epoch GMOS images from B15 are indicated with 1$\sigma$ error ellipses (yellow) while the centroid of the bulge component in each filter is indicated with a white cross. The weak North-South gradient and the diffraction spike to the West in the F160W image is caused by a bright star to the North of the visible field, while the object to the North-East of both images is an unrelated galaxy. The position angle of the XShooter slit is also indicated on the image, with the slit itself centred on the host galaxy.}
\label{fig:HST_cutouts}
\end{center}
\end{figure}

The resulting images are shown in Figure \ref{fig:HST_cutouts}. In each, it is clear that the host has a highly complex, irregular morphology that is broadly split into two main components: one a compact, bulge-like component and the other a more diffuse, extended component which could constitute a disc. Within the extended component there exists a star-forming complex or ``knot'' on the western edge of the host complex that is most evident in the F606W imaging (see Figure \ref{fig:HST_cutouts}). Given that this ``knot'' of emission is located far from the position of the transient, it likely has little bearing on the nature of this event. For the purposes of the analysis (and in part due to the findings of the analysis of the X-Shooter spectrum in subsection \ref{Xshooter}) we consider the photometry of the two main components separately as well as that of the whole system. From this point forwards, parts of the host's morphology will be referred to separately as the bulge and the extended component (the bulk of the diffuse component including the star-forming knot).

In the F160W imaging, photometry of the entire system was completed in an aperture that covered the entire host complex (radius 1.6$^{\prime\prime}$). In order to separate the bulge and extended components, two methods were used. The first method assumes that the extended and bulge components are spatially separable such that the extended component contributes minimally to a small aperture (radius 0.4$^{\prime\prime}$) centred on the bulge component. The extended component photometry is then simply the difference in flux between the total and bulge aperture photometry.

The second method is based on modelling completed using {\sc galfit} \citep[][]{Peng2002,Peng2010} using 2 S\'{e}rsic profiles, one for each component. Due to the low surface brightness and large extent of the extended component, coupled with the existence of a number of nearby features that were masked from the fit (nearby unrelated galaxy, diffraction spike from bright star etc.), the fit is unable to simultaneously constrain the half-light radius and uncertainties. Instead, the fit is carried out with the half-light radius fixed to the best fit value (1.3$^{\prime\prime}$). The result is a wide, very flat (S\'{e}rsic index $\ll 0.1$) extended disc, while the bulge component is much more compact (half-light radius of 0.22$\pm$0.04$^{\prime\prime}$, S\'{e}rsic index of 2.1$\pm$0.7).  A star-forming knot of emission to the west of the host (as seen in Figure \ref{fig:HST_cutouts}) was masked out to accommodate the fit. In order to determine its contribution to the extended component, aperture photometry was used on the model subtracted image centred on the knot emission (radius 0.5$^{\prime\prime}$).

In the F606W imaging, the morphology of the host is too complex to fit with simple radial profiles and thus photometry is determined entirely through integrated flux with apertures matched to those from the F160W imaging described above.

The resultant photometry is detailed in Table \ref{tab:PhotomJ1112-2}. Aperture photometry was corrected for aperture losses assuming the sources were point-like, which provides a minimum aperture correction. In practice, for the large apertures considered these corrections are minimal. All photometry has been corrected for Galactic extinction with E(B$-$V) = 0.253 $\pm$ 0.009\footnote{based on values derived from \citet[][]{Schlafly2011} accessed via the NASA/IPAC Infrared Science Archive http://irsa.ipac.caltech.edu/applications/DUST/}, assuming the Fitzpatrick extinction law \citep[][]{Fitzpatrick1999}.

\begin{table}
\begin{center}
\begin{tabular}{c c c c}
\hline
{\bf Filter} 	& {\bf Component} 	& {\bf Aperture} 	& {\bf Modelled} 	\\
 				& 					& {\bf Magnitude} 	& {\bf Magnitude}	\\ [0.5ex]
\hline 
160W 			& 	Total 			& 	22.17$\pm$0.07 	&   22.11$\pm$0.08	\\
 	 			& 	Bulge 			& 	23.48$\pm$0.10	&   23.56$\pm$0.13	\\
 	 			& 	Extended 		& 	22.56$\pm$0.11	&   22.53$\pm$0.04	\\
606W				&	Total  			& 	23.09$\pm$0.11	&	--	\\	
				& 	Bulge 			& 	25.08$\pm$0.11 	&   -- 				\\
	 			& 	Extended  		& 	23.28$\pm$0.14 	& 	--	\\

\hline
\end{tabular}
\caption[{\em HST} photometry of the host of Swift J1112-8238, broken down by component]{{\em HST} photometry of the host of Swift J1112-8238. The component refers to the portion of the complex morphology being analysed, namely the total system, the position of the IR bulge and the extended remainder of the system. In the F606W imaging, the photometry for both the 1.1$^{\prime\prime}$ visible emission aperture and 1.6$^{\prime\prime}$ F160W matched aperture cases are shown. Where applicable, aperture and Galactic extinction corrections have been applied. The total photometry for the F160W modelled magnitude comes from the sum of the contributions of the bulge and extended modelled photometry plus the additional contribution of the ``knot'' of emission (which was excluded from the fit) determined in aperture photometry of the model subtracted image. Note that the total magnitude of the system is consistent in the modelled and aperture methods.}
\label{tab:PhotomJ1112-2}
\medskip
\end{center}
\end{table}

\subsection{X-Shooter Spectroscopy}\label{Xshooter}

Spectroscopy of the host of Swift J1112-8238 was taken using X-Shooter \citep[][]{Vernet2011} on the Very Large Telescope (VLT) on 2014 December 19 (MJD 57010) beginning at 04:44{\sc ut} and on 2014 December 2015 20 (MJD 57011) beginning at 04:43{\sc ut}. In each observation the UVB and VIS arms were exposed for a total of 2720\,s (4$\times$680) while the NIR arm exposure time was 2400\,s (4$\times$600). The data was taken in NOD mode (ABBA pointing) with 5$^{\prime\prime}$ offsets between the A and B nod positions in the spatial direction (i.e. along the slit).

Since 2012 August, the atmospheric dispersion correctors (ADCs) mounted on the UVB and VIS arms of X-Shooter have been offline due to malfunction, usually requiring observations to be oriented along the parallactic angle to reduce losses. Unfortunately, the presence of the aforementioned bright star due north of our source made this impossible and forced a position angle of $\mytilde50^{\circ}$. Instead, wide slits (1.6$^{\prime\prime}$ and 1.5$^{\prime\prime}$ in the VIS and UVB arms respectively) and maximised exposure times were used in the UVB and VIS arms to attempt to counter this effect. Given our previous inferred redshift based on lower resolution spectroscopy (as discussed in B15), strong emission lines were not expected to be observable in the UVB arm or the blue end of the VIS arm where the effect is at its worst.

The data was reduced both via the standard {\sc reflex} pipeline \citep[][]{Freudling2013} and independently via the method of \citet[][]{Modigliani2010}, producing consistent results. Flux calibration was completed with respect to the spectrophotometric standard star Feige110. Telluric corrections were determined through use of the telluric star Hip058859, but were not found to be relevant in the wavelength range of interest.

\begin{figure}
\includegraphics[width=8.4cm]{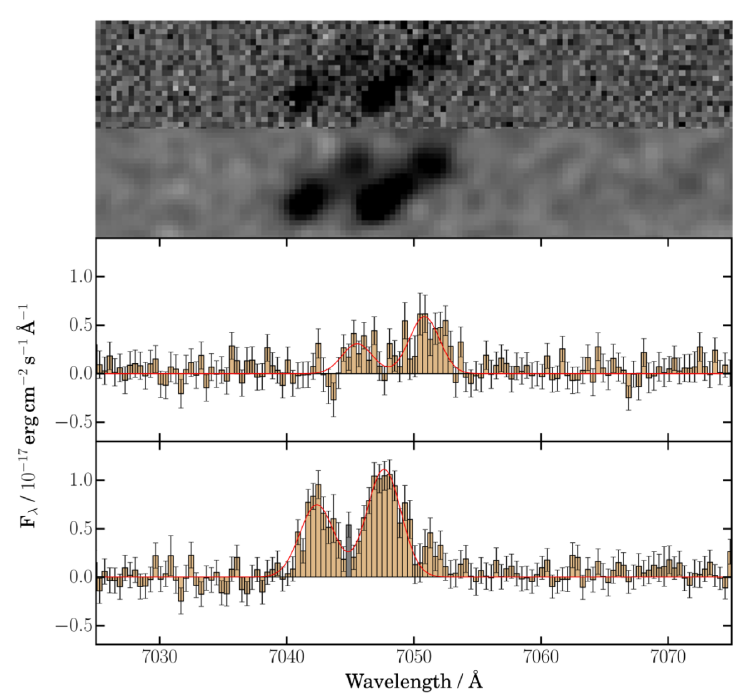}
 \caption[The X-Shooter spectrum centred on the {[}O{\sc ii}{]}$\lambda\lambda3726,3729$ emission doublet]{The X-Shooter spectrum centred on the feature at $\mytilde7045$\AA\, interpreted as the [O{\sc ii}]$\lambda\lambda3726,3729$ emission doublet. (Top Panel) The reduced 2D X-Shooter spectrum and (Second Panel) the same spectrum with a 3 pixel Gaussian smoothing. The doublet nature of the line is clearly visible, as is the existence of two spatially resolved components with a clear offset in velocity space. (Third Panel) The northern and (Bottom Panel) southern components extracted as 1D spectra with 1$\sigma$ errorbars. The red line indicates the Gaussian line profiles fitted to the data.}
\label{fig:OIISpec}
\end{figure}

There was insufficient flux to produce a detection of the continuum level the unbinned data at any wavelength, though faint continuum was detected upon extreme binning of the VIS arm data. A strong, doublet emission feature, coincident with the unresolved line at $\mytilde7045$\AA\ from the GMOS and FORS2 spectra presented in B15, was clearly visible. Interestingly, the line is split both spatially and in velocity space into two clear components as can be seen in Figure \ref{fig:OIISpec}, which is consistent with an asymmetry observed in the GMOS spectrum from B15. Other emission lines were also visible at wavelengths of $\mytilde9465$\AA, $\mytilde9190$\AA\ and $\mytilde8200$\AA\ in the VIS arm and at $\mytilde12405$\AA\ in the NIR, all displaying the same two component nature except in the case of the $\mytilde$9465\AA\ line, where strong sky line emission masked the second component. No features were detected in the UVB arm. All of the detected features are consistent with the original interpretation of the single line discussed in B15 as being the [OII] emission doublet, coincident as they are with the relative positions of H$\gamma$, H$\beta$, [O{\sc iii}]$\lambda5007$ and H$\alpha$. The [O{\sc iii}]$\lambda4959$ line was covered by sky line emission for both spatially resolved components, while the [O{\sc iii}]$\lambda5007$ line was obscured for the southern component. Limits were also placed on the [N{\sc ii}]$\lambda6583$ line emission.

One dimensional spectra of each of the two spatially resolved components were extracted and each line of sufficient significance was fitted with a gaussian profile. Based on the position angle of the slit, we determine that the two components (upper and lower) constitute separate detections of the bulge and extended/western knot components of the host system respectively. This results in measured redshifts of 0.8904$\pm$0.0001 for the bulge and 0.8895$\pm$0.0001 for the extended component, a velocity offset of 303$\pm$47\,km\,s$^{-1}$ and giving an overall redshift for the complex of 0.8900$\pm$0.0005, consistent with that determined in B15. The line widths were deconvolved with the instrumental resolution to determine the velocity dispersion of the lines, 105$\pm$15km\,s$^{-1}$ and 120$\pm$10km\,s$^{-1}$ for the bulge and extended components respectively.

The line fluxes were determined separately for each component and corrected for Galactic extinction. Due to the complex morphology of the host and inability to spatially resolve the continuum level of the separate components, no attempt was made at determining slit losses and thus the quoted fluxes are systematically underestimated by as much as a factor of a few. The line fluxes are presented in Table \ref{tab:SpecResults}.

\begin{table}
\begin{center}
\begin{tabular}{c c c c}
\hline
{\bf Line}     	& {\bf Observed}		& {\bf Bulge}   & {\bf Extended} \\
{\bf Identity}     	& {\bf Wavelength (\AA)}		&    &  \\
\hline \relax
[O{\sc ii}]$\lambda$3727 	& 7046		& 0.76$\pm$0.17  & 2.20$\pm$0.19\\ \relax
[O{\sc ii}]$\lambda$3729 	& 7051		&  1.50$\pm$0.23  & 3.35$\pm$0.19 \\  \relax
H$\alpha$          			& 12409		&  2.27$\pm$0.24  & 3.68$\pm$0.36 \\
H$\beta$          			& 9192		&  0.59$\pm$0.14  & 1.13$\pm$0.14 \\
H$\gamma$         			& 8207		&  0.32$\pm$0.13  & 0.47$\pm$0.12 \\ \relax
[O{\sc iii}]$\lambda$5007 	& 9467		&  1.20$\pm$0.18  & -- \\ \relax
[N{\sc ii}]$\lambda$6583 	& N/A		&	$<0.72$		& $<1.02$ \\
\hline
\end{tabular}
\caption[The emission line fluxes determined from the X-Shooter spectrum of the host of Swift J1112-8238]{Line fluxes of the host of Swift J1112-8238 determined from the XShooter spectrum. The fluxes are determined for each component separately, while the observed wavelength is quoted only for the bulge component. Fluxes are quoted in units of 10$^{-17}$erg\,cm$^{-2}$\,s$^{-1}$, have been corrected for Galactic extinction but have no correction for slit losses. Limits on [N{\sc ii}]$\lambda$6583 emission are given to 3$\sigma$. The [O{\sc iii}]$\lambda$5007 extended component was contaminated by strong sky emission and was not recoverable}
\label{tab:SpecResults}
\medskip
\end{center}
\end{table}

\subsection{Late-time Radio Observations of the Host}

Radio flares were observed in both of the previous candidates \citep[][]{Bloom2011,Cenko2012,Zauderer2013}. The well-observed radio lightcurve of Swift J1644+57 indicates that these events are capable of producing bright, long-term radio emission years after the initial event \citep[][]{Zauderer2013}. This radio emission originates from the jet colliding with the circunumclear gas and producing synchrotron emission and was predicted, by analogy to GRB afterglows, from TDEs prior to J1644+57 in \citet[][]{Giannios2011}.

Observations of Swift J1112-8238 were undertaken using the Australia Telescope Compact Array (ATCA), with an initial epoch on 2015 Jan 30 ($\mytilde1300$ days post trigger in the observer frame)\footnote{Observations associated with programme C3002, PI: Stanway}. The observations were taken simultaneously in two bands, centred at 5.5 and 9.0\,GHz. The telescope was in its most elongated 6A configuration, with baselines between 5.938 and 0.337\,km aligned East-West, and Earth rotation synthesis was used to improve coverage of the $uv$-plane. A total on-source integration of 95 minutes was divided over hour angles spanning nearly 12 hours, allowing good reconstruction of the synthesised beam which had a $2.5^{\prime\prime}\times1.9^{\prime\prime}$ full-width at half-maximum at 5.5\,GHz. 

Secondary phase calibrations were performed using regular observations of PKS\,1057-797, and absolute flux and bandpass calibration were determined through observations of PKS\,1934-638 (the standard calibrator for ATCA). The data was flagged for radio frequency interference, calibrated, imaged and deconvolved with the synthesised beam using the standard software package {\sc miriad} \citep{Sault1995}. Each band comprised 2048 channels, each of 1\,MHz bandwidth. Multi-frequency synthesis images were constructed using natural weighting and the full bandwidth between the flagged edges of each band.

\begin{figure}
\includegraphics[width=8.4cm]{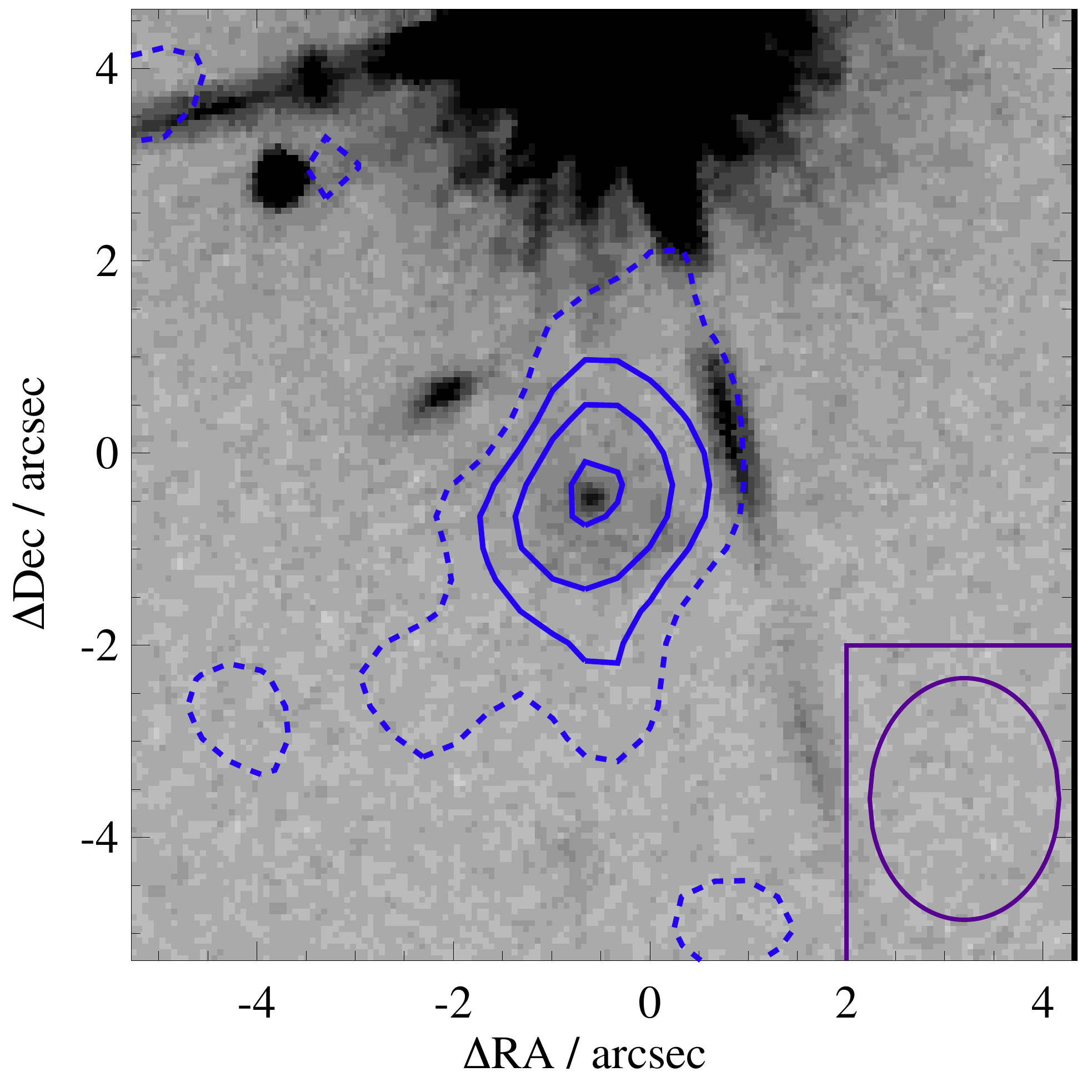}
\caption[The early epoch radio emission contours overlayed on the F160W {\em HST} image of Swift J1112-8238]{The early epoch 5.5\,GHz radio contours overplotted on the F160W {\em HST} image. The radio beam width is also represented in the lower right corner. The solid contours represent 2, 3 and 4 times the sky RMS while the dashed contours represent 1$\sigma$ fluctuations. While alignment between optical and radio images is difficult, the emission appears to be centred on the bulge component.}
\label{fig:Radiooverlay}
\end{figure}

In both bands, a faint source was identified coincident with the coordinates of Swift J1112-8238 (RA=11:11:47.6 Dec=-82:38:44.44 in the 5.5\,GHz imaging with a positional uncertainty of $\mytilde0.25^{\prime\prime}$; see Figure \ref{fig:Radiooverlay}). Photometry was completed by fitting point sources to the emission (see Table \ref{tab:RadioFlux}). Attempts to fit an extended source to the 5.5\,GHz data yields a flux estimate consistent within the point source estimate but with significant uncertainty on both the source size and resultant integrated flux, suggesting that the signal to noise ratio is insufficient to perform such an analysis.

Further observations were made with an identical instrumental configuration through 2016 May 11-16, approximately 1.3 years later in the observer frame\footnote{Observations associated with programme C3099, PI: Brown} with a total exposure time of 8.5 hours. The data was reduced following the same procedure. Fluxes are shown in Table \ref{tab:RadioFlux}. This second epoch of observations again revealed a point source in the 5.5\,GHz band with a flux consistent with the previous epoch. However, the source had apparently declined in the 9.0\,GHz band to a non-detection with a 3$\sigma$ upper limit of 54$\mu$Jy. 

\subsubsection{Swift J2058+05}
During the 2016 May observing run, comparison observations were also completed on the second rTDF candidate Swift J2058+05 on 2016 May 14 with a total exposure time of 3.75 hours. Secondary phase calibration was achieved through observations of PKS 2121+053 and  PKS\,1934-638 was used for primary flux calibration. The data were reduced in {\sc Miriad} as described above.

Fluxes for J2058+05 are also shown in Table \ref{tab:RadioFlux}. Somewhat surprisingly given the greater distance to the event ($z=1.12$ cf $z=0.89$ for Swift J1112-8238), a source was well detected at the position of Swift J2058+05 in both bands. The northern position of the source and uv-coverage of the observation meant that the beam was strongly elongated along the north-south direction (FWHM 34$^{\prime\prime} \times 1.1^{\prime\prime}$). However, the lack of nearby sources in optical images makes confusion unlikely.

\begin{table*}
\begin{center}
\begin{tabular}{c c c c c}
\hline
{\bf Source}     & {\bf Observation} & {\bf MJD}   & {\bf 5.5\,GHz Flux} & {\bf 9.0\,GHz Flux} \\
{\bf Name}		&	{\bf Date}		&	& {\bf /$\mu$Jy} & {\bf /$\mu$Jy} \\
\hline
Swift J1112-8238 & 2015 Jan 30 & 57052	& 76$\pm$15 & 70$\pm$29 \\
				 & 2016 May 11-16 & 57519-57524 & 70$\pm$11 & $<54$ (3$\sigma$) \\
Swift J2058+05 & 2016 May 14 & 57522 & 225$\pm$15 & 236$\pm$13 \\
\hline
\end{tabular}
\caption[The radio fluxes determined from the ATCA observations of Swift J1112-8238 and Swift J2058+05]{The observed radio flux from the ATCA observations. All photometry is based on fitting point sources to the images.}
\label{tab:RadioFlux}
\medskip
\end{center}
\end{table*}

\section{Discussion}\label{sec:discussion}

\subsection{Host Morphology and Transient Position}\label{Morph}

In both bands of the {\em HST} imaging the host of Swift J1112-8238 shows clear evidence of a complex morphology with at least two main components. The first is a simple bulge-like structure visible in both the F160W and F606W imaging. In the F160W imaging, modelling shows the bulge is clearly extended with a half-light radius of 0.22$\pm$0.04$^{\prime\prime}$. At this redshift, this constitutes a physical size of 1.7$\pm$0.3\,kpc. The S\'{e}rsic index of $\mytilde$2.1 is also consistent with typical values for galactic bulges, placing it on the boundary between a classical bulge (or elliptical galaxy) and a disk-like bulge (pseudobulge) \citep[][]{Fisher2008,Gadotti2009}, though the large error on this value cannot distinguish between the two. 

The bulge component of the F606W imaging is marginally more extended than point sources in the field, with FWHM 0.11$^{\prime\prime}$ compared to 0.08$^{\prime\prime}$ for other point sources based on measurements using the {\sc iraf} \citep[][]{Tody1986,Tody1993} function {\sc imexam}. Assuming gaussian profiles for both the point source function and the bulge morphology, the half-light radius of the bulge is $\mytilde0.04^{\prime\prime}$ = 0.3\,kpc. If true, this indicates the F606W emission comes primarily from a small region within the IR bulge component. We also note that the bulge centroids in each waveband are consistent to within measurement errors ($\mytilde$0.02$^{\prime\prime}$), which could suggest that even late after the outburst there is still some contribution from the transient.

The second component is much more diffuse, with modelling of the F160W imaging suggesting the presence of a wide flat disc with half-light radius of $\mytilde$1.3$^{\prime\prime}$=10\,kpc. This component displays considerable inhomogeneity with the most obvious feature being the knot of emission most visible in the F606W imaging on the western edge of the system, most-likely a star-forming complex.

This morphology is also apparent in the X-Shooter spectrum of the host, with two clear components separated in both the spatial and dispersion directions, where the offset in velocity is $\mytilde$300 kms$^{-1}$. The two sources may then be separate galaxies that are in the process of a tidal interaction or merger, which would also explain the irregular nature of the second component. However we cannot rule out the presence of a bulge embedded in a disc. The bulge and extended components have quite different colours, the bulge exhibiting substantially more infrared emission than the bluer knot with F606W-F160W ($\mytilde$V-H) colours of 1.60$\pm$0.15 and 0.72$\pm$0.18 for the bulge and extended component respectively. This indicates very different stellar populations in the two regions.

The position of the transient with respect to this newly discovered complex morphology is shown in Figure \ref{fig:HST_cutouts}. Note that the error is dominated by the poor seeing and signal-to-noise of the original ground-based transient images with the matching to the {\em HST} imaging contributing minimally to the final uncertainty. The transient is clearly associated with the bulge as opposed to the extended component. The transient position is between 1$\sigma$ and 2.5$\sigma$ from the centroid position of the IR and optical bulge, depending on the GMOS observation and {\em HST} band used. Combining the probability distributions of the two independently determined positions, the best position of the transient places it at an angular distance of 0.14$\pm$0.06$^{\prime\prime}$ which corresponds to a projected physical distance of 1.1$\pm$0.5\,kpc. Thus the transient position remains consistent with the centroid of the host, albeit somewhat loosely, and thus a supermassive black hole origin remains plausible.

\subsection{Internal Extinction, Metallicity and Classification of the Host}

Numerous transient phenomena, including supernovae \citep[e.g. ][]{Filippenko1997} and tidal disruption flares \citep[][]{Arcavi2014}, are known to produce transient line emission. However, the nature of the origin of these lines typically leads to extremely wide profiles, with widths of many hundreds to thousands of km\,s$^{-1}$, something that is clearly not the case with $\mytilde100$\,km\,s$^{-1}$ widths of these lines. A number of tidal disruption flares have also possibly been identified through transient line emission alone, such as those found in the galaxies SDSS J095209.56+214313.3 \citep[][]{Komossa2008a} and SDSS J074820.67+471214.3 \citep[][]{Wang2011a}. In these cases, the line emission can be considerably narrower, more akin to that observed here, and that fade over years to decades. However, these events also showed strong coronal line emission, such as high ionisation iron lines that are not present in this case. Given this, and the lack of evidence for line emission in the suspected related rTDF candidates \citep[e.g.][]{Pasham2015}, the following analysis is made under the assumption that the observed line emission comes from the host alone. Further late-time spectroscopy could confirm this in the future. 

The internal extinction of the bulge and extended components were determined through their Balmer decrements based on the theoretical values for Case B recombination of a gas with temperature of 10$^{4}$\,K and electron density of 10$^{2}$\,cm$^{-3}$, as is commonly used in the literature \citep[$H_{\alpha}/H_{\beta} = 2.86$;][]{Osterbrock2006}. This results in $E(B-V)_{\mathrm{line}} = 0.27\pm0.18$ and 0.12$\pm$0.12 respectively. Extinction corrections are applied throughout the following analysis assuming the dust extinction law of \citet[][]{Calzetti2000}.

The procedures of \citet[][]{Kewley2008} were used to determine the metallicity of the host. The calibrations of \citet[][]{McGaugh1991}, \citet[][]{Kobulnicky2004}, \citet[][]{Zaritsky1994} and \citet[][]{Pilyugin2001} were applied to the Galactic and internal extinction corrected line fluxes of the bulge component of the host. The lack of an [O{\sc iii}]$\lambda5007$ detection (or the [O{\sc iii}]$\lambda4959$ line) due to coincident strong sky line emission precludes a similar analysis of the extended component. With one exception, we find the calibrations are consistent with a metallicity of $12 + \mathrm{log_{10}(O/H)}\,\mytilde\,$8.5$\pm$0.2, that is a metallicity slightly sub-solar \citep[8.69,][]{Asplund2009}. The exception, the \citet[][]{Pilyugin2001} calibration, is determined via the electron temperature (T$_e$) method, a method known to break down at metallicities of $\mytilde8.4$ \citep[e.g.][]{Brown2016}. 

\begin{figure}
\includegraphics[width=8.4cm]{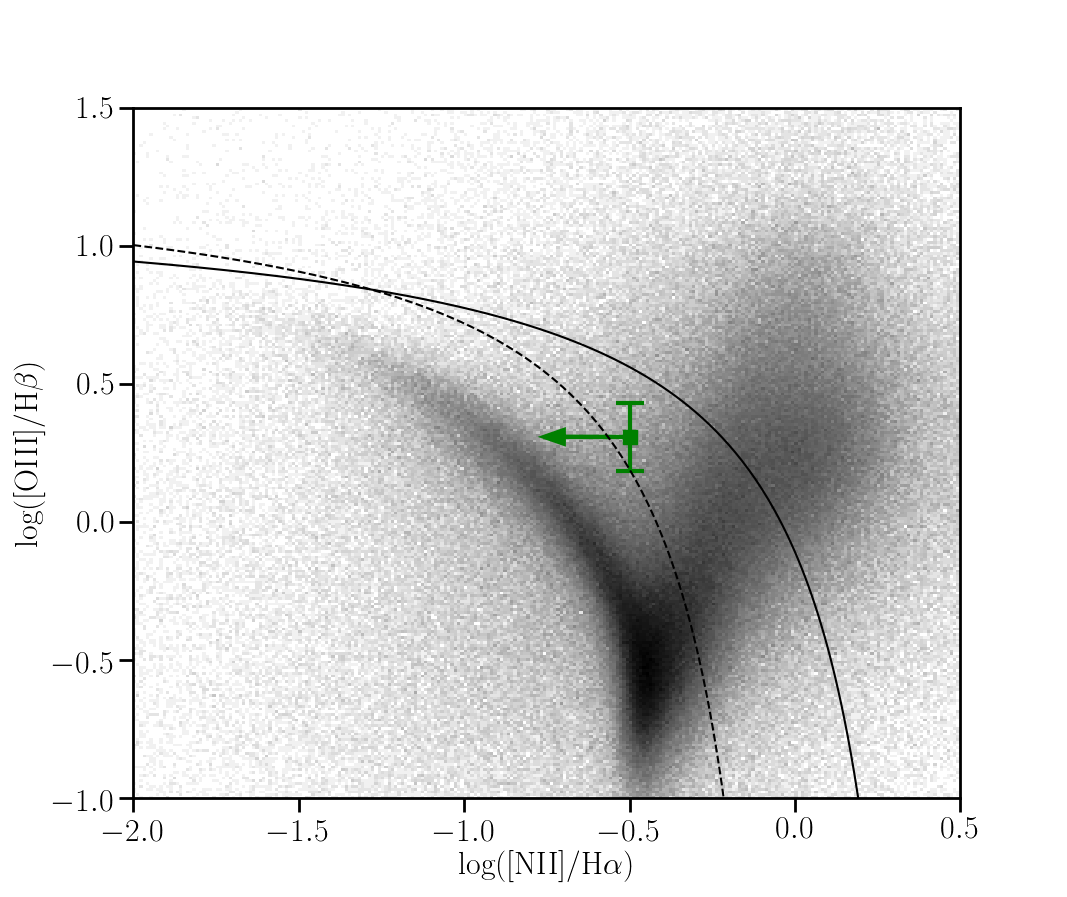}
 \caption[The Baldwin Phillips Terlevich (BPT) diagram of the bulge component of the host of Swift J1112-8238]{The BPT diagram of the bulge component of the host of Swift J1112-8238 plotted as a 3$\sigma$ upper limit on the [NII]/H$\alpha$ line ratio. The MPA-JHU analysed sample of SDSS DR8 galaxies \citep[][]{Brinchmann2004,Tremonti2004} are plotted in grayscale. The extremal \citep[][]{Kewley2001} and more conservative \citep[][]{Kauffmann2003b} delineations between the AGN and star-forming regions of the diagram are plotted as solid and dashed lines respectively. The upper limit places the host well within the extremal delineation and is strongly suggestive of belonging to the star-forming locus.}
\label{fig:BPTdiagram}
\end{figure}

A Baldwin, Phillips \& Terlevich (BPT) diagram \citep[][]{Baldwin1981} can distinguish between a star-forming and AGN-dominated classification for the host. The resulting diagram is plotted in Figure \ref{fig:BPTdiagram} along with a representative sample of SDSS galaxies and the dividing lines between the two regions of the plot as determined in \citet[][]{Kewley2001} and \citet[][]{Kauffmann2003b}. The delineation determined in \citet[][]{Kewley2001} is based on the extremal case that is designed to include all starburst galaxies at the expense of including some AGN, as opposed to the more conservative delineation determined in \citet[][]{Kauffmann2003b}. Obscuring sky lines at the position of [O{\sc iii}]$\lambda5007$ in the extended spectrum makes the determination of constraining limits impossible. The upper limit on [N{\sc ii}]/H$_{\alpha}$, while formally consistent with a composite interpretation, is indicative of a star-formation driven radiative field and makes it unlikely that there is strong AGN activity in the host. The known, but poorly constrained, evolution of the BPT diagram further pushes the host into the star forming locus \citep[][]{Kewley2013}.

\subsection{Stellar Mass and Black Hole Mass}

At a redshift of 0.89, the F160W filter represents a rest frame wavelength of approximately 8000\AA. This wavelength, longward of the Balmer break, can be used to estimate the stellar mass of the host under the assumption of a flat spectral energy distribution (SED). 

A calibration, based on the K-band absolute magnitude, is provided by \citet[][]{Savaglio2009}. Using the observed F160W absolute magnitudes corrected for internal extinction yields stellar mass estimates for the bulge and knot components of ($1.4\pm0.2)\times$10$^{9}$\,M$_{\odot}$ and $(2.9\pm0.3)\times10^{9}$\,M$_{\odot}$ respectively. All stellar and black hole masses are subject to an additional 0.1\,dex systematic uncertainty from the internal extinction correction not included in the quoted uncertainties and there is a further $\mytilde50\%$ uncertainty from the \citet[][]{Savaglio2009} calibration.

Unfortunately, a K-correction term ($2.5\log_{10}\left(1+z\right)$) was inadvertently applied in the wrong direction in B15; the corrected absolute magnitudes of the flare and host in the $i^{\prime}$ band are $M_{\mathrm{flare}} = -20.0$ and $M_{\mathrm{host}} = -20.3$. This corresponds to a host mass determination of $4\times10^{8}\,$M$_{\odot}$ based on the \citet[][]{Kauffmann2003b} rest $g^{\prime}$-band (observed $i^{\prime}$-band) mass to light ratios and an inferred SMBH mass of $\mytilde2\times10^{6}\,$M$_{\odot}$ from the scaling relation of \citet[][]{Bennert2011}. 

The \citet[][]{Savaglio2009} calibration gives a mass that is approximately an order of magnitude larger than the $g^{\prime}$-band inferred stellar mass estimate. Both of these calibrations are associated with a large intrinsic scatter and neither probes the rest-frame infrared in this source. It is also worth noting that these calibrations are based on the average of large populations and may not be appropriate for implementation in an unusual interacting system.

It is also possible to improve our estimate of the SMBH mass in this system. {\em HST} imaging has revealed the presence of a possible bulge-disk system. Applying the bulge mass scaling relation of \citet[][]{Haring2004} results in a mass of (1.3$\pm$0.2)$\times10^{6}$\,M$_{\odot}$, though with few low mass black holes studied in \citet[][]{Haring2004} it is unclear how accurate this value is. Under the assumption that each component (bulge and extended) instead represents a separate galaxy, we estimate central SMBH masses of $(6.8\pm0.1)\times10^6$M$_{\odot}$ and $(1.6\pm0.3)\times10^7$M$_{\odot}$ respectively based on the scaling relation of \citet[][]{Bennert2011}. Finally, the estimates produced from the bulge mass to black hole mass relation of \citet[][]{Kormendy2013} results in values of $(3.3\pm0.5)\times10^{6}$M$_{\odot}$ and $(7.8\pm1.0)\times10^{6}$M$_{\odot}$ (with an additional 0.28\,dex scatter not included here).  Each estimate is well within the $10^{8}$M$_{\odot}$ limit for the disruption of a Sun-like star \citep[][]{Rees1988}.

\subsection{Star Formation Rate and Stellar Population}

We use the F606W absolute magnitude ($\mytilde3100$\,\AA\,rest) as a star formation rate indicator. \citet[][]{Moustakas2006} derive a $U$-band ($\mytilde3600$\,\AA) conversion of $(1.8\pm1.0)\times10^{-43}$\,L(U)\,$M_{\odot}$\,yr$^{-1}($erg\,s$^{-1})^{-1}$. This implies SFRs of 0.16$\pm$0.09(0.11)\,M$_{\odot}$\,yr$^{-1}$ and 0.54$\pm$0.31(0.20)M$_{\odot}$\,yr$^{-1}$ for the bulge and extended components. The quoted uncertainty on each result comes primarily from the large uncertainty in the calibration, a result of the reddening variation that dominated the sample the relation was derived from. We indicate a further systematic uncertainty due to the internal extinction correction in brackets. This can be very large, as much as 100\% at rest-frame ultraviolet wavelengths. 

Alternate star formation rates can be derived from certain line luminosities. Both H$\alpha$ and [O{\sc ii}] have been used for this, subject to uncertainties in metallicity, dust extinction and nebular gas conditions. We derive star formation rates of 1.06$\pm$0.12(0.72) and 1.01$\pm$0.09(0.49)\,$M_\odot$\,yr$^{-1}$ for the bulge and extended components from H$\alpha$ \citep[using the calibration of][]{Murphy2011} and 1.6$\pm$0.4(1.3) and 1.6$\pm$0.5(1.8)\,M$_\odot$\,yr$^{-1}$ from [O{\sc ii}] \citep[using the calibration of][]{Kewley2004}\footnote{The \citet[][]{Murphy2011} calibrated values have been corrected from the \citet[][]{Kroupa2001} initial mass function to the \citet[][]{Salpeter1955} initial mass function assumed in the \citet[][]{Moustakas2006} and \citet[][]{Kewley2004} calibrations}. These are broadly consistent, although the line indicators suggest a higher star formation rate in the bulge component than the continuum. This may indicate that continuum extinction has been underestimated or alternatively that the physical conditions differ from those assumed in the calibrations (100 Myr old continuously star forming stellar population). The total star formation within this system is constrained to be less than 3.2$\pm$0.6\,M$_\odot$\,yr$^{-1}$.

Although the two components have similar star formation rates, their very different morphologies are reflected in quite different line properties. As Figure \ref{fig:OIISpec} makes clear their line ratios, including between components of the strong [O\,{\sc II}] emission doublet, differ suggesting a difference in the nebular gas properties or the ionizing stellar population.

\subsection{Properties of the Radio Emission}

If interpreted as star formation, and accounting for the radio spectral slope, the rest frame 1.4GHz radio luminosity implied by the first epoch of radio observations suggests a star formation rate of a few hundred solar masses per year \citep[][]{Condon2002}. The spectral slope is, in reality, poorly constrained. This very high implied star formation rate, is 2 orders of magnitude higher than even the largest estimates from the UV and optical measures. The moderate internal extinction determined in this case makes dust obscured star formation unlikely to be the cause of the discrepancy. Further, the radio spectral index ($\alpha$ defined as $S_{\nu} \propto \nu^{\alpha}$, where $S_{\nu}$ is the flux per unit frequency, $\nu$) in both epochs is higher than would be expected for star formation associated synchrotron emission, which tends to be within the range $-0.5<\alpha<-1$ \citep[e.g.][]{Condon1992,Thompson2006,Seymour2008}. Coupled with the apparent evolution in flux and spectral index, the radio emission is inconsistent with coming solely from star formation.

The radio emission appears to be coincident with the bulge component (Figure \ref{fig:Radiooverlay}). This is further evidence for an association with the transient flare, a finding that is also backed up by the possible evolution of the source. An alternative explanation is that it is is due to unrelated AGN activity from the bulge's central supermassive black hole. However, the narrow emission lines visible in the X-Shooter spectrum and their line ratios suggest this is not the case.

\begin{figure}
\includegraphics[width=8.4cm]{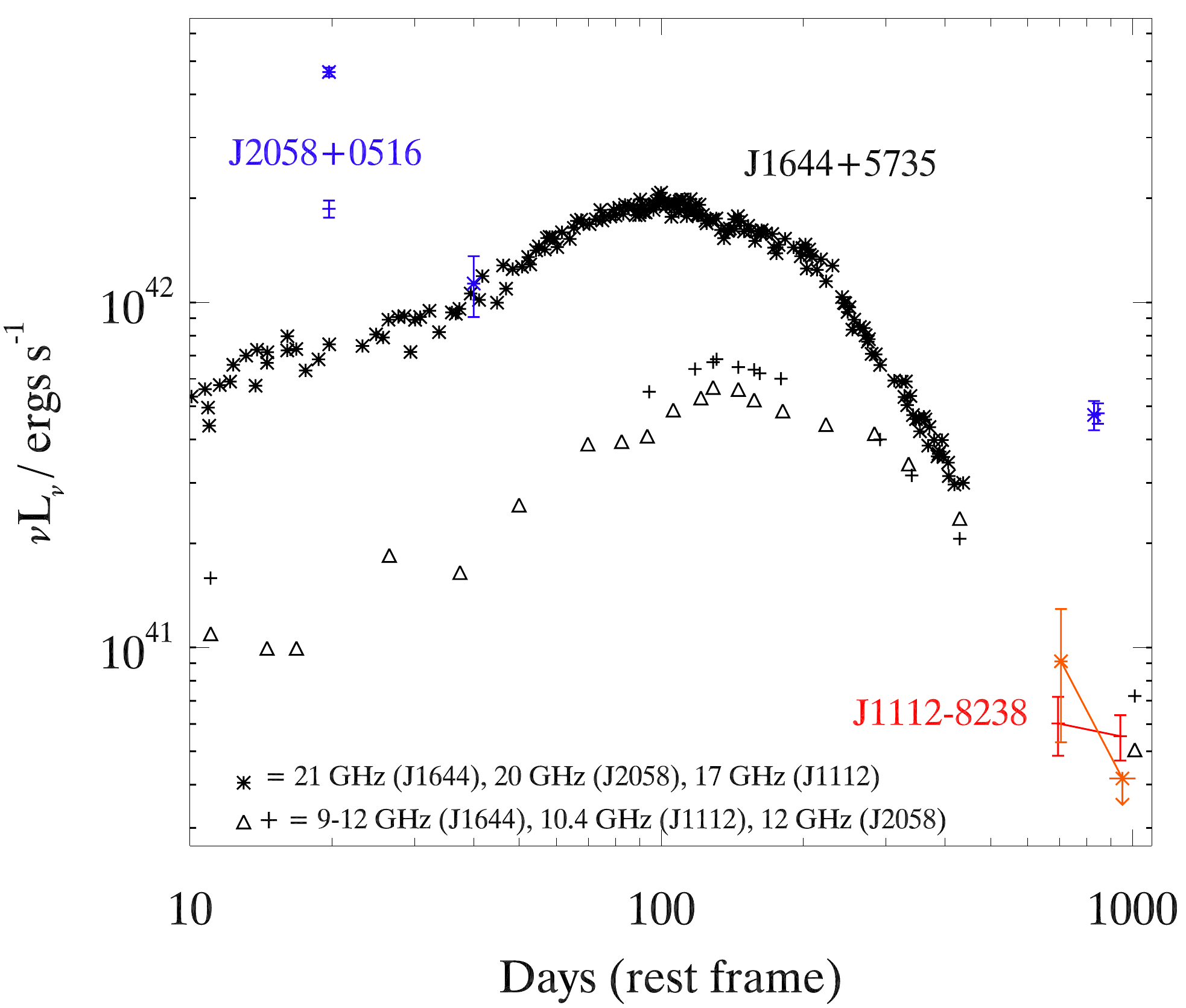}
 \caption[The radio lightcurves of the three relativistic tidal disruption flare candidates]{The radio lightcurves of the three rTDF candidates plotted in rest-frame time and scaled to the redshift of Swift J1644+57, the best studied of the three rTDF candidates. The lightcurves are plotted for Swift J1644+57 (black), Swift J2058+05 (blue) and Swift J1112-8238 (red and orange). The lightcurves are plotted at frequencies of $\mytilde10$ and $\mytilde20$\,GHz rest-frame compiled from \citet[][]{Zauderer2011}, \citet[][]{Berger2012}, \citet[][]{Zauderer2013} and from a VLA observation (associated with VLA/14A-423 PI: Zauderer) for Swift J1644+57, while the Swift J2058+05 observations are from \citet[][]{Cenko2012} and \citet[][]{Pasham2015}. The luminosities of the three flares are consistent to within an order of magnitude at late times. Interestingly, with the exception of the $\mytilde$40 day Swift J2058+05 observation, which unlike the other observations was made using Very Long Baseline Interferometry, the lightcurves are remarkably consistent with scaled versions of the Swift J1644+57 lightcurve.}
\label{fig:RadioLightcurve}
\end{figure}

\begin{figure}
\includegraphics[width=8.4cm]{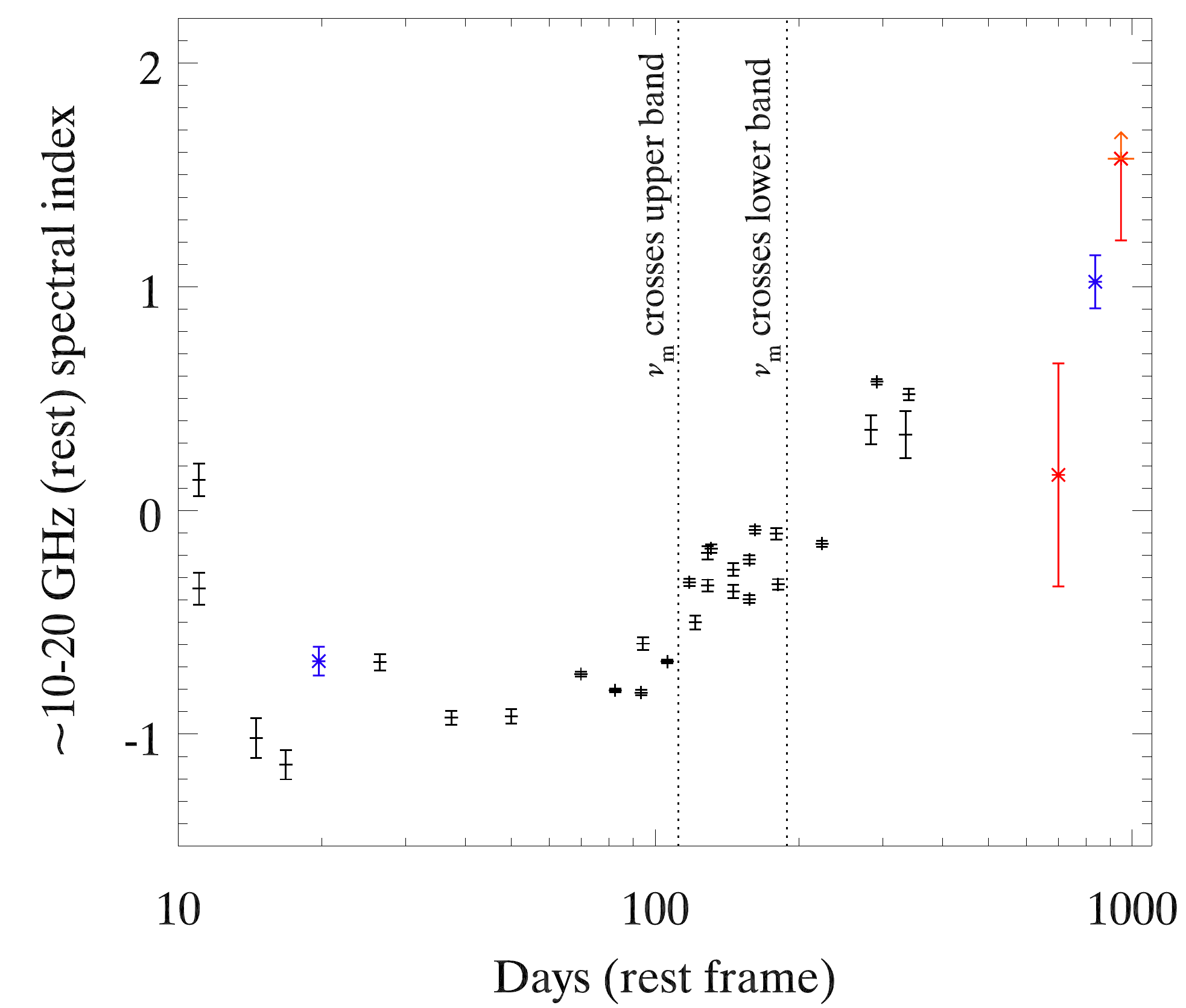}
 \caption[The spectral index evolution of the three relativistic tidal disruption flare candidates]{The spectral index evolution of the three rTDF candidates, Swift J1644+57 (black), Swift J2058+05 (blue) and Swift J1112-8238 (red). The data for Swift J1644+57 is compiled from contemporaneous observations from \citet[][]{Zauderer2011}, \citet[][]{Berger2012} and \citet[][]{Zauderer2013}, while the Swift J2058+05 observations are from \citet[][]{Cenko2012} and \citet[][]{Pasham2015}. The vertical dashed lines indicate the times when the peak frequency of the modelled emission passes through the observed bands \citep[][]{Metzger2012}. Note that, while the uncertainities on the Swift J1112-8238 observations are large, all three candidates are consistent with following the same spectral evolution.}
\label{fig:RadioAlpha}
\end{figure}

As such, it is plausible that the emission is instead directly associated with the transient, possibly an rTDF. Unfortunately, as mentioned in B15, the archival radio limits are not deep enough to determine whether these observations are brighter than the quiescent level of the host. However, it is possible to compare them to the previous rTDF candidates. In Figure \ref{fig:RadioLightcurve}, the radio lightcurves of all three rTDF candidates are plotted in their rest frame. We plot published data for the best studied burst, Swift J1644+57, chosen to approximate the rest frequencies of our observations and for comparison we reduced a single late time epoch of archival observations taken at the VLA at 21 and 9\,GHz (PI: Zauderer). The late time emission of all three candidates are within an order of magnitude of each other at the same epoch. Further, the evolution of Swift J1112-8238 is consistent with the shallow rest-frame 10\,GHz and steep 20\,GHz evolution of Swift J1644+57. The lightcurve of Swift J2058+05 is, at first glance, quite different to the other two, with an apparent sharp decline in flux by the rest-frame 40 day observation. This measurement was made through VLBA observations which may not be directly comparable with the other observations. Given that at X-ray and $\gamma$-ray wavelengths Swift J1112-8238 and Swift J1644+57 had similar luminosities while Swift J2058+05 was somewhat more luminous, it would not be unreasonable to expect a similar behaviour in radio luminosity. With the exception of the anomalous VLBA point, that appears to be the case.

As shown in \citet{Genezerov2017}, we expect that the peak radio luminosity of the radio emission to scale appropximately as E$_\mathrm{jet}^{0.6}$ and with external gas density as $n_\mathrm{gas}^{1.2}$, so the factor $\sim4$ in radio luminosity for Swift J2058+05 compared to J1644+57 may indicate a jet energy higher by a factor of $\sim$10 or a circumnuclear density higher by a factor of $\sim$3.  Given the variation we expect in the latter, the observed variation between events is physically plausible. By contrast, and as was the case for X-ray flux, this suggests that  Swift J1112-8238 had a very similar jet-energy to Swift J1644+57. 

The late time radio emission from TDFs is expected to be virtually isotropic \citep{Mimica2015} and thus off-axis observations of this class of relativistic TDFs are clearly distinct from those of observed thermal TDEs, whose radio fluxes at comparable times are significantly lower \citep[][]{vanVelzen2013,Bower2013}. However the sparse sampling precludes further interpretation, emphasising the need for further monitoring of members of this class.

The spectral evolution of all three candidates (where contemporaneous observations in two or more bands allow) also shows a striking similarity (see Figure \ref{fig:RadioAlpha}). Both Swift J1644+57 and Swift J2058+05 begin with a negative spectral index that evolves as expected towards synchrotron self-absorption at late times \citep[$\alpha\mytilde1$,][]{Metzger2012}. The uncertainties on the spectral indices of the observations of Swift J1112-8238 are large, but are nonetheless consistent with the same spectral evolution.

\section{Implications for the interpretation of the flare}

Having obtained high resolution imaging, we confirm that the transient centroid is likely to be associated with one component of an interacting system. Within this, it is consistent with, but not precisely aligned to, the most massive and most compact emission region. This offset would not be unheard of in events identified as tidal disruption flares, with PTF10nuj and PTF11glr \citep[][]{Arcavi2014}, and ASASSN15oi \citep[][]{Holoien2016b} each had offsets of varying significance from their host's centres.

However, it remains possible that the flare is not associated with accretion onto a supermassive black hole but instead with an unusual core collapse event origin. Such a core collapse event could result from a burst of star formation triggered by the merging system as molecular gas clouds are subjected to shocks and tidal effects \citep[e.g.][]{Bournaud2011}. Merger triggered star formation commonly occurs in one of two places - in the nuclear region of the merging galaxy \citep[][]{Keel1985} or on the boundary \citep[as seen in the Antenna Galaxy;][]{Wang2004}. The latter scenario may explain the star-forming complex seen on the south-western edge of the system in the F606W imaging, but this is inconsistent with the transient location. A possibly related class of event, the Ultra-long GRB \citep[ULGRBs; see e.g.][]{Levan2014}, has been associated with massive core collapse through the detection of a supernova following the initial burst of a single event \citep[][]{Greiner2015}. As discussed in B15, the properties of the ULGRBs and the three rTDF candidates differ in that ULGRBs have much shorter gamma-ray flare durations, as defined by the T$_{\mathrm{90}}$ measure (10$^4$ seconds for the ultra-long GRBs compared with the 10$^6\,$s for the Swift J1644+57-like events), and far less luminous late-time X-ray emission.

The detection of an optical/NIR rebrightening in the long-term evolution of Swift J1644+57 \citep[][]{Levan2016} with an absolute magnitude, color and duration consistent with that of a superluminous supernova, could suggest all of the detected extreme duration gamma-ray flares originate from massive star collapse. However, this rebrightening has a number of alternative possible explanations including the reverberation of X-ray emission similar to the effect used to map the central regions of AGN, based on a possible time lag between the X-ray and optical lightcurves \citep[][]{Yoon2015}. It could also be explained in terms of a late peaking component of thermal or synchrotron emission associated with a tidal disruption flare \citep[][]{Levan2016}. With an absolute magnitude of $M_{i^{\prime}} = -20.0$, Swift J1112-8238 is comparable in luminosity to typical superluminous supernovae. However, the peak optical observations made $\mytilde20$ days post trigger ($\mytilde10$ days rest-frame) place it on a somewhat shorter timescale than the tens of days rise times seen in SLSNe \citep[e.g.][]{GalYam2012}. Unfortunately, the sparsely sampled optical lightcurve of Swift J1112-8238 makes determining the presence or absence of an underlying supernova impossible in this case.

Recent work on the hosts of hydrogen-poor superluminous supernovae shows a preference for a narrow range of properties. If Swift J1112-8238 were associated with a superluminous supernova, it would have an atypical host galaxy. The bulge component of the host is consistent with being amongst the more massive and more star-forming of the superluminous supernova hosts \citep[][]{Lunnan2014,Angus2016,Perley2016}. Meanwhile, its near solar metallicity disfavours its inclusion with the hosts of hydrogen-poor SLSNe \citep[][]{Leloudas2015}. However the much wider range of galaxies playing host to the hydrogen-rich variant, with metallicities up to 8.9, makes it difficult to rule out a superluminous supernova association.

By comparison, the interacting nature of the host has some interesting parallels in other TDF hosts. Studies such as those by \citet[][]{Arcavi2014} and \citet[][]{French2016,French2016a} have shown that a disproportionate fraction of TDF hosts show signs of being members of a rare class of galaxy known as the E+A galaxies. These galaxies are thought to be the result of a relatively recent and significant burst of star formation \citep[][]{Quintero2004} that may have been triggered by a merger event. They make up less than 1\% of local galaxies, and yet a majority of thermal TDF hosts show E+A-like properties such as strong Balmer absorption lines with minimal ongoing star formation. The reason for the greater rates of TDFs within these galaxies is likely due to the presence of enhanced stellar number densities in the cores of these galaxies which increases the number of stars able to be funnelled into the loss cone \citep[][]{Stone2016a}. However, E+A galaxies are typically defined as having no observable H$\alpha$ and [O{\sc ii}] emission \citep[][]{Goto2007,Yamauchi2008} and it appears unlikely that the host of Swift J1112-8238 has the very low H$\alpha$ equivalent width found for typical TDF hosts by \citet[][]{French2016}. This suggests that it is a rather younger star forming system, perhaps only recently undergoing its merger event.

The question then is at what point in the evolution of a merging/post-merger system does the enhanced stellar number density occur. It is possible that some degree of over-density could present itself while the merger is still on going, as may have been seen here. The reason then for the relative lack of ongoing examples in thermal TDFs is that the merger timescale of only a $\mytilde$1\,Gyr \citep[e.g.][]{Cox2008} is shorter than the lifetimes of Sun-like stars ($\mytilde10$\,Gyr). However, this ratio is not so large as to exclude the chance of seeing a reasonable fraction in ongoing mergers as may have happened here.

Alternatively, the merging nature of the host opens other channels for the enhanced tidal disruption flare rates. Tidal interactions perturb stellar orbits and can enhance the rate of TDFs by as much as two orders of magnitude. This effect, while dependent on the mass of each galaxy, occurs when the galaxies come within a few effective radii of one another \citep[][]{Liu2009}. At the same time, migration of the black holes towards the dynamical centre provides an additional boost. This may also lead to a spatial offset between a TDF location and the apparent host centroid (as may be seen in Swift J1112-8238) since the black holes are not guaranteed to be coincident with the bulk of the stellar mass. Finally, the merger of the supermassive black holes is also expected to greatly enhance the rate of tidal disruption flares for a short period of time. In the few decades following the SMBH merger, rate of tidal disruption flares may increase to $\mytilde$0.1\,yr$^{-1}$ \citep[][]{Stone2011}. While the short duration of this enhancement makes this effect unlikely to be a large contributor to the global TDF rate, it is nonetheless expected to be observed on occasion in galaxies undergoing, or shortly following, a merger. Indeed because the dynamical friction timescale for the infall of SMBHs is only of order 1\,Gyr \citep[e.g.][]{Just2011}, similar to the galaxy merger timescale, it might be possible to observe the effects of the SMBH merger while the galaxies are still heavily distorted. In any case, the concurrence of SMBH mergers and high tidal disruption flares rates is of great interest in the current age of multi-messenger astronomy.

In both of the previous rTDF candidates, a strong diagnostic of their nature was the detection of a rising radio flare with properties that indicated the production of a moderately relativistic jet \citep[][]{Zauderer2011,Berger2012,Cenko2012,Pasham2015}. The radio observations outlined in this work confirm the existence of a similar feature in the late-time evolution of Swift J1112-8238. Given the very low optical and UV inferred star formation rates and lack of any other AGN indicators, there is no other clear explanation for the radio emission. The possibility of similar lightcurves and spectral evolution between the three rTDF candidates is also tantalisingly suggestive of a connection. However, the lack of long-term monitoring in two of the candidates, and the low significance of the detections of Swift J1112-8238, preclude more definite inferences. This highlights the need for regular follow-up of candidates with radio observations to build up an understanding of the radio evolution of these events. 

One outstanding question surrounding the origin of these events is the apparent lack of new candidates. All three firm candidates were detected within three months of one another in the first half of 2011. While previously suggested to be little more than a statistical fluke, the longer the dearth of candidates continues, the more puzzling it becomes. In recent months, however, new possible candidates have been detected. These have included Fermi J1544-0649/Swift 154419.7-064915/ASASSN17gs \citep[][]{Ciprini2017}, an X-ray transient detected with {\em Fermi} and {\em Swift} that may also be associated with an optical transient detected by the ASASSN supernova survey. A further event detected by {\em Swift}, GRB170714A \citep[][]{DAi2017}, is a long-lived gamma-ray transient that also displays some similarities to the rTDF candidates. However, GRB170714A has a much shorter duration in gamma-rays and the lack of detection in radio with the VLA \citep[][]{Horesh2017} and NOEMA \citep[][]{DeUgartePostigo2017} indicates this burst is instead a ULGRB, while Fermi J1544-0649 may instead belong to a set of BL Lac type objects with similar, albeit fainter, flares \citep[][]{Kawase2017}. As such, for the moment it seems that the three rTDF candidates remain the only events of their type and the lack of similarity with other events means the most likely explanation for their temporal coincidence is still an odd statistical fluke. Only future observations of candidate flares will be able to confirm this.

\section{Summary}

Our results may be summarised as follows:

\begin{enumerate}

\item{X-Shooter spectroscopy has confirmed the previously determined redshift of the candidate as $z=0.8900\pm0.0005$.}
\item{Both X-Shooter spectroscopy and {\em HST} imaging show the host has a complex morphology, consistent with an interaction or merger.}
\item{The transient is loosely associated with the bulge component centroid, although formally offset by $2.2\sigma$.}
\item{The detection of radio emission from the host of Swift J1112-8238 likely comes from the transient flare. Comparison of the radio lightcurves and spectral evolution of all three rTDF candidates suggest they may be consistent with a single evolutionary behaviour.}
\item{All available evidence remains consistent with a tidal disruption flare origin for Swift J1112-8238.}
\end{enumerate}

Based on these findings, it seems likely that the three rTDF candidates do indeed share a common origin, although better sampling of the evolution of these sourecs would have strengthened this conclusion. Priority must therefore be placed on obtaining systematic follow-up of future candidates.

\section*{Acknowledgements}
GCB thanks the Midlands Physics Alliance for a PhD studentship and the Institute of Advanced Study, Warwick, for postdoctoral research funding. AJL and ERS acknowledge funding from the UK Science and Technology Facilities Council (STFC) associated with grant number ST/L000733/1. TK acknowledges support through the Sofja Kovalevskaja Award to PS from the Alexander von Humboldt Foundation of Germany. We also acknowledge travel funding from the STFC grant ST/M006492/1, and Royal Astronomical Society travel support. Based on observations made with ESO Telescopes at the La Silla Paranal Observatory under programme ID 094.B-0703. Based on observations made with the NASA/ESA Hubble Space Telescope, obtained at the Space Telescope Science Institute, which is operated by the Association of Universities for Research in Astronomy, Inc., under NASA contract NAS 5-26555. These observations are associated with program 13869. The Australia Telescope Compact Array is part of the Australia Telescope National Facility which is funded by the Australian government for operation as a national facility managed by CSIRO. The authors made use of the York Extinction Solver \citep[YES,][]{McCall2004}. The authors made use of Ned Wright's Javascript Cosmology Calculator \citep[][]{Wright2006}.




\bibliographystyle{mnras}
\bibliography{/home/gregbrown/Documents/library} 








\bsp	
\label{lastpage}
\end{document}